\documentstyle[12pt,epsf]{article}

\baselineskip=\normalbaselineskip

\ifx\TwoupWrites\UnDeFiNeD\else\target{\magstepminus1}{11.3in}{8.27in}
        \source{\magstep0}{7.5in}{11.69in}\fi
\newfont{\fourteencp}{cmcsc10 scaled\magstep2}
\newfont{\titlefont}{cmbx10 scaled\magstep3}
\newfont{\authorfont}{cmcsc10 scaled\magstep1}
\newfont{\fourteenmib}{cmmib10 scaled\magstep2}
        \skewchar\fourteenmib='177
\newfont{\elevenmib}{cmmib10 scaled\magstephalf}
        \skewchar\elevenmib='177
\makeatletter
\newcommand\nonsequentialeqnum{
        \@addtoreset{equation}{section}
        \def\theequation{\arabic{section}.\arabic{equation}}}
\newif\ifp@bblock  \p@bblocktrue
\newcommand\nopubblock{\p@bblockfalse}
\newcommand\topspace{\hrule height 0pt depth 0pt \vskip}
\newcommand\p@bblock{\begingroup \tabskip=\hsize minus \hsize
        \baselineskip=1.5\ht\strutbox \topspace-2\baselineskip
        \halign to\hsize{\strut ##\hfil\tabskip=0pt\crcr
        \the\Pubnum\crcr\the\date\crcr}\endgroup}

\renewcommand\titlepage{\ifx\TwoupWrites\UnDeFiNeD\null\vspace{-1.7cm}\fi
\vskip0.6cm
        \ifp@bblock\p@bblock \else\hrule height 0pt \relax \fi}
\makeatother
\newtoks\date
\newtoks\Pubnum
\newtoks\pubnum
\Pubnum={
YITP-96-43 \crcr
KUNS1413 \crcr
HE(TH)96/14\crcr
hep-th/9609210
}
\date={\today}
\newcommand{\frontpageskip}{\vspace{12pt plus .5fil minus 2pt}}
\renewcommand{\title}[1]{\frontpageskip
        \begin{center}{\titlefont #1}\end{center}\par}
\renewcommand{\author}[1]{\frontpageskip\par\begin{center}
        {\authorfont #1}\end{center}
        \nobreak
        }

\newcommand{\address}[1]{\par\begin{center}{\sl #1}\end{center}\par}

\renewcommand{\thanks}[1]{\footnote{#1}}
\renewcommand{\abstract}{\par\frontpageskip\centerline{
        \fourteencp Abstract}
        \vspace{8pt plus 3pt minus 3pt}}
\begin{document}

\pubnum{93-37}
%\date{September 1996 \crcr (Revised)}
\date{September 1996}
\titlepage

\renewcommand{\thefootnote}{\fnsymbol{footnote}}
\title{
Nonperturbative Effects in Noncritical Strings\\
\vskip5pt
with Soliton Backgrounds
%\thanks{}
}

\author{
Masafumi Fukuma${}^{1\,}$\thanks{
e-mail address: {\tt fukuma@yukawa.kyoto-u.ac.jp}}
and Shigeaki Yahikozawa${}^{2\,}$\thanks{
e-mail address: {\tt yahiko@gauge.scphys.kyoto-u.ac.jp}}
}

\address{
${}^1$
Yukawa Institute for Theoretical Physics\\
Kyoto University, Kyoto 606-01, Japan \\
{}~\\
${}^2$
Department of Physics\\
Kyoto University, Kyoto 606-01, Japan \\
}

\newcommand{\bc}{\begin{center}}
\newcommand{\ec}{\end{center}}
\newcommand{\bl}{\begin{flushleft}}
\newcommand{\el}{\end{flushleft}}
\newcommand{\bi}{\begin{itemize}\begin{enumerate}}
\newcommand{\ei}{\end{enumerate}\end{itemize}}
\newcommand{\bt}{\begin{tabbing}}
\newcommand{\et}{\end{tabbing}}
\newcommand{\np}{\newpage}
\newcommand{\pset}{\setcounter{page}{1}}
\newcommand{\wh}[1]{w^{({#1})}}

\renewcommand{\thefootnote}{\arabic{footnote}}
\setcounter{footnote}{0}
\newcommand{\cleqn}{\setcounter{equation}{0} \indent}
\renewcommand{\theequation}{\thesection.\arabic{equation}}
\newcommand{\beqa}{\begin{eqnarray}}
\newcommand{\eeqa}{\end{eqnarray}}
\newcommand{\n}{\nonumber}
\newcommand{\nn}{\nonumber \\ }
\newcommand{\eq}[1]{(\ref{#1})}
\newcommand{\bC}{{\bf C}}
\newcommand{\cD}{{\cal D}}
\newcommand{\cH}{{\cal H}}
\newcommand{\cO}{{\cal O}}
\newcommand{\Psid}{\Psi^\dagger}
\newcommand{\norm}[1]{{\parallel {#1} \parallel}^2}
\newcommand{\nnorm}[1]{{{\parallel {#1} \parallel}^{\prime\,2}_l}}
\newcommand{\del}{\partial}
\newcommand{\db}{{\bar{\delta}}}
\newcommand{\gbar}{{\bar{g}}}
\newcommand{\dl}
           {\left[\,\frac{dl}{\,l\,}\,\right]}
\newcommand{\Det}{\,\mbox{Det}\,}
\newcommand{\Tr}{\,\mbox{Tr}\,}
\newcommand{\ldot}{\dot{l}}
\newcommand{\const}{\mbox{const.\ }}
\newcommand{\bra}[1]{\left\langle\,{#1}\,\right|}
\newcommand{\ket}[1]{\left|\,{#1}\,\right\rangle}
\newcommand{\vev}[1]{\left\langle\,{#1}\,\right\rangle}
\newcommand{\bZ}{{\rm {\bf Z}}}
\newcommand{\svac}{\bra{\sigma}}
\newcommand{\pvac}{\ket{\Phi}}
\newcommand{\rphi}{\varphi}
\newcommand{\rphih}{\hat{\rphi}}
\newcommand{\dphi}{\del\rphi}
\newcommand{\dphih}{\del\hat{\rphi}}
\newcommand{\cb}{\bar{c}}
\newcommand{\gint}{\oint^{\,p}}
\newcommand{\dds}{\frac{d s}{2\pi i}}
\newcommand{\ddz}{\frac{d\zeta}{2\pi i}}

%%%%%%%%%%%%%%%%%%%%%%%%%%%%%%%%%%%%%%%%%%%%%%%%%%%%%%%%%%%%%%%%%
% Abstract
%%%%%%%%%%%%%%%%%%%%%%%%%%%%%%%%%%%%%%%%%%%%%%%%%%%%%%%%%%%%%%%%%
\begin{abstract}

We explicitly construct soliton operators in $D<2$ (or $c<1$)
string theory,
and show that the Schwinger-Dyson equations
allow solutions with these solitons as backgrounds.
The dominant contributions from 1-soliton background are
explicitly evaluated in the weak coupling limit, and
shown to agree with the nonperturbative analysis of string equations.
We suggest that fermions should be treated as fundamental
dynamical variables since both macroscopic loops and solitons
are constructed in their bilinear forms. \\

\end{abstract}

%%%%%%%%%%%%%%%%%%%%%%%%%%%%%%%%%%%%%%%%%%%%%%%
% section 1
%%%%%%%%%%%%%%%%%%%%%%%%%%%%%%%%%%%%%%%%%%%%%%%
\section{\protect\large{\protect\bf Introduction}}
\cleqn

Solitons in string theory have played important roles in
understanding the nonperturbative dynamics of strings \cite{m-soliton}.
Such solitons in the weak coupling region
are expected to have the typical dependence
on the string coupling constant $g$.
In fact, in the string perturbation theory,
perturbation series diverge as $(2h)!$ (not $h!$) with genus $h$,
and thus the leading nonperturbative effect is
generally expected to have the form $e^{-{\rm const}\,/g}$ \cite{s}.
D-branes (especially D-instantons) \cite{p1} actually have
this behavior, since they are defined
as the contribution from the (Dirichlet-)boundaries of world sheets,
and the leading contribution comes from a disk amplitude of
order $g^{-1}$, which will be exponentiated through a special
combinatorics in space-time picture \cite{p2-g}.
This observation, however, makes difficult the constructive approach
to closed string field theory with simple covariant actions,
since if some action exists and yields genus expansion
with weight $g^2$ for each loop (genus),
then natural nonperturbative effects will have the form
$e^{-{\rm const}\,/g^2}$.
One possibility to solve this problem is,
that there are more fundamental dynamical variables
which describe both elementary strings and these solitons
in a unified manner.

In the present letter, we point out that this possibility is actually
realized in $D<2$ string theory,
which can be constructed as the double scaling limits \cite{bk-ds-gm}
of matrix models.
In particular, we identify the fundamental dynamical variables
from which both elementary strings and solitons
are constructed as their bilinears.
We further evaluate the nonperturbative effects
due to 1-soliton background, and show that they
have the form $e^{-{\rm const}\,/g}$ with the correct exponent,
which are consistent with the nonperturbative analysis of string
equations.
{}Further investigation of the dynamics for these variables
will be reported in our future communication.

We start our discussion with recapitulating the Schwinger-Dyson
equation approach to noncritical string theory \cite{d0,fkn1,dvv},
aiming to fix our notation.
{}For strings in $D<2$, or equivalently 2D quantum gravity
coupled to some conformal fields with central charge $c<1$,
string variables should be invariant for reparametrizations
of the world sheet.
Such variables that appear naturally in matrix models are
the {\em macroscopic-loop} operators \cite{bdss}.
Their connected $N$-point correlation functions $v(l_1,\cdots,l_N;g)$
are defined as the sum over fluctuating (connected) surfaces
with $N$ boundaries of length $l_1,\cdots,l_N$
along which spins of conformal matters have the same state.
{}For later use, it is convenient to introduce their Laplace transforms,
which have an asymptotic expansion in genus $h$ of the form
\beqa
  w(\zeta_1,\cdots,\zeta_N;\,g)&\equiv&\int_0^\infty dl_1 \cdots dl_N
        e^{-\zeta_1 l_1-\cdots-\zeta_N l_N} v(l_1,\cdots,l_N;\,g)\n\\
  &=& \sum_{h\geq 0}\,g^{-2+2h+N}\,\wh{h}(\zeta_1,\cdots,\zeta_N).
\eeqa
Here $g$ is the renormalized string coupling constant.

If the conformal field theory is the minimal $(p,q)$ model
with central charge $c=1-6(p-q)^2/pq$,
then $w(\zeta_1,\cdots,\zeta_N;\,g)$ has the following form of
Laurent expansion around $\zeta=\infty$:
\beqa
  w(\zeta_1,\cdots,\zeta_N;\,g)\,=\,\frac{1}{p^N}
        \sum_{n_1\cdots n_N}\,\zeta_1^{-n_1/p-1}\cdots
        \zeta_N^{-n_N/p-1}\,
        w_{n_1,\cdots,n_N}(g).\label{1.2}
\eeqa
{}For positive $n_i$, the coefficient $w_{n_1,\cdots,n_N}(g)$
is identified with the connected correlation function of
the {\em microscopic-loop} operators $\cO_{n_i}$:
$\,\vev{\cO_{n_1}\cdots\cO_{n_N}}_c\,$
$=\,$ $\sum_{h\geq 0}g^{-2+2h+N}\,
\vev{\cO_{n_1}\cdots\cO_{n_N}}^{(h)}_c$.
These operators $\cO_n$ $(n=1,2,3,\cdots)$ correspond to
the physical operators in the Liouville gravity \cite{pk-kpz-ddk}
obtained as the surface integral of the gravitationally-dressed,
spinless primary fields with (undressed) conformal dimension
$\Delta^{(0)}_{r,s}\,=\,[(qr-ps)^2-(p-q)^2]/4pq$
and $n=|qr-ps|$.

{}For this $(p,q)$ case, any correlation function including
$\cO_n$ with $n$ multiple of $p$ always vanishes.
{}Furthermore, solving the Schwinger-Dyson equations of
the matrix models with appropriate double scaling limit,
we find infinitely many relations among the correlation functions.
They are compactly expressed as the $W_p$ constraint \cite{fkn1,dvv}
on the exponential of the generating function $F(j)$
of connected correlation functions:
\beqa
  {}F(j)&=&\vev{ \exp\left\{
        \sum_{n\geq1;\,n\not\equiv 0~({\rm mod}\,p)}
        j_n\,\cO_n\right\}}_c \n\\
        &=&\log \left\langle 0 \left|
        \exp\left\{\sum_{n\geq1;\,n\not\equiv0\,({\rm mod}\,p)}
        \left(j_n-B_n/g\right)\alpha_n\right\}
        \right| \Phi \right\rangle.
\eeqa
Here $\alpha_n~(n\geq1)$ are the positive part of
the bosonic oscillators $\alpha_n~(n\in\bZ)$ with commutation
relation $[\alpha_n,\,\alpha_m]\,=\,n\,\delta_{n+m,\,0}$, and
$\bra{0}$ is the vacuum satisfying $\bra{0}\alpha_n=0 ~ (n\leq0)$.
The state $\ket{\Phi}$ satisfies the vacuum condition of the
$W_p$ algebra made from the $\bZ_p$-twisted bosons
$\dphih_k(\zeta) = (1/p)\sum_{n\equiv k\,({\rm mod}\,p)}
\zeta^{-n/p-1}\alpha_n$ $(k=1,2,\cdots,p-1)$,
whose explicit form is not important here and can be found in refs.\
\cite{fkn1,dvv,fkn2}.
The $B_n$'s are backgrounds which characterize the theory,
and the $(p,q)$ case is realized by $B_n \,=\,
(B_1,B_2,\cdots,B_{p+q},0,0,\cdots)$ with nonvanishing $B_{p+q}$.
With the help of string equations \cite{md},
it can be further shown \cite{fkn2,fkn3} that
$\tau(x)=\bra{0}\exp\{ \sum_{n\geq1}x_n\alpha_n \}\ket{\Phi}$ is
a $\tau$ function of the $p^{\rm th}$ reduced KP hierarchy,
and also that the $W_p$ constraint is automatically enhanced
to the $W_{1+\infty}$ constraint;
We will fully use the latter fact later,
giving an explicit basis of the $W_{1+\infty}$ algebra \cite{winf}.
We have found that the string coupling constant $g$ can be placed
in the generating function in this simple manner.
It is easy to check that this form actually reproduces correct
genus expansion of the Schwinger-Dyson equations.

{}For one- and two-point functions ($N=1,\,2$), however,
the right-hand side of \eq{1.2} includes the terms
with negative ${n_i}$ of order $g^{-1}$ and $g^0$, respectively.
(Note that both come from spherical topology.)
Such terms are difficult to be identified with
the correlation functions of microscopic-loop operators,
and (misleadingly) termed ``non-universal parts''
although they do scale properly in the continuum limit.
They are found to be \cite{fkn1,gn,mss,iikmns}
\beqa
  w_{\rm non}(\zeta)&=&-\,\frac{1}{pg}\,\sum_{n}\,n B_n \zeta^{n/p-1}
        \label{1.4}\\
  w_{\rm non}(\zeta_1,\zeta_2)&=&
        \frac{1}{p^2\,(\zeta_1\zeta_2)^{1-1/p}\,
        (\zeta_1^{1/p}-\zeta_2^{1/p})^2}
        \,-\,\frac{1}{(\zeta_1-\zeta_2)^2} \n\\
  &=&\del_{\zeta_2}\left[
        \frac{\frac{1}{p}\,\sum_{i=0}^{p-1}
        \left(\zeta_2/\zeta_1\right)^{i/p}-1}
        {\zeta_1-\zeta_2}\right]. \label{1.5}
\eeqa
Since it is those macroscopic-loop operators that are
obtained naturally in the scaling limit of matrix models (see, 
{\em e.g.}, \cite{mss} for further investigation on this point),
we will regard the expression including the non-universal terms
as being fundamental.

This paper is organized as follows.
In section 2 we first introduce the twist operator $\sigma(\zeta)$
to ``untwist" the twisted bosons,
and then construct free fermions
with which the Schwinger-Dyson equations are most simply expressed.
We identify the string field that describes macroscopic loops.
In section 3, we first construct soliton fields from the fermions,
and show that the Schwinger-Dyson equations have solutions
with backgrounds of these solitons.
We then make an explicit calculation of the nonperturbative effects
for 1-soliton background and show that they agree with the value
appearing in the nonperturbative analysis of string equations.
Section 4 will be devoted to conclusions with some speculations.

%%%%%%%%%%%%%%%%%%%%%%%%%%%%%%%%%%%%%%%%%%%%%%%%%%%%%%%%
% section 2
%%%%%%%%%%%%%%%%%%%%%%%%%%%%%%%%%%%%%%%%%%%%%%%%%%%%%%%%
\section{\protect\large{\protect\bf Fermionic Representation of the
Schwinger-Dyson
Equations with Twist Operator}}
\cleqn

In order to simplify the discussion on the $W_{1+\infty}$ constraint,
we first introduce a twist operator $\sigma(\zeta)$ \cite{dfms-br}
which enables us to deal with $\dphih_k(\zeta)$
as being untwisted.
The twist operator $\sigma(\zeta)$ thus has the property
that $\dphih_k(\zeta)$ has a definite monodromy when going around
the point at which $\sigma(\zeta)$ is inserted.
This is ensured by the following OPE for the operators
$\dphih_k(\zeta)~(k=0,1,\cdots,p-1)$ and $\sigma(\zeta)$:
\beqa
  \dphih_k(\zeta)\,\dphih_l(\zeta')&\sim&
        \frac{g_{k\,l}/p}{(\zeta-\zeta')^2} \\
  \dphih_k(\zeta)\,\sigma(\zeta')&=&
        \frac{1}{p}\sum_{m\leq -1}(\zeta-\zeta')^{-m-k/p-1}\,
        (\alpha_{mp+k}\sigma)(\zeta'),
\eeqa
where $g_{k\,l}\equiv\delta_{k+l\equiv0\,({\rm mod}\,p)}$.
We have introduced a monodromy-free boson $\dphih_0(\zeta)$,
which is necessary to reproduce the correct non-universal terms
in later discussions.
By denoting the $SL(2,\bC)$ invariant vacuum by $\bra{{\rm vac}}$,
the state $\bra{0}$ in the preceding section is now interpreted
as the state $\bra{\sigma}\equiv\bra{{\rm vac}}\sigma(\infty)$,
and $\dphih_k(\zeta)$ has a monodromy $\omega^{-k}$
around $\zeta=\infty$:
\beqa
  \svac\,\dphih_k(e^{2\pi i}\zeta)\,=\,
        \omega^{-k}\svac\,\dphih_k(\zeta).
\eeqa

To express the $W_{1+\infty}$ algebra, we introduce
``orthonormal basis'' $\dphi_a(\zeta)$ $(a=0,1,\cdots,p-1)$
from the monodromy-diagonalizing basis $\dphih_k(\zeta)$ by
\beqa
  \dphi_a(\zeta)&=&
        \sum_{k=0}^{p-1}\,\omega^{-ka}\,\dphih_k(\zeta)~~~~~
  \left(\,\Leftrightarrow\,\,\dphih_k(\zeta)\,=\,
        \frac{1}{p}\sum_{a=0}^{p-1}\,\omega^{ka}\,
        \dphi_a(\zeta)\,\right)\\
  \dphi_a(\zeta)\,\dphi_b(\zeta')
        &\sim&\frac{\delta_{a\,b}}{(\zeta-\zeta')^2},
\eeqa
which has the following monodromy around $\svac$:
\beqa
  \svac\,\dphi_a(e^{2\pi i}\zeta)&=&\svac\,\dphi_{[a+1]}(\zeta)
        {}~~~~([a]\equiv a~({\rm mod}\,p)).
\eeqa
The $W_{1+\infty}$ currents $W^k(\zeta)$ $(k=1,2,\cdots)$ are then
written \cite{fkn2} as
\beqa
  W^k(\zeta)\,=\,-\,\frac{1}{k}\,
	\sum_{a=0}^{p-1}:e^{\rphi_a(\zeta)}\,
        \del_\zeta^k\,e^{-\rphi_a(\zeta)}: .
\eeqa
Note that since we make a diagonal sum over $a$,
there is no monodromy for $W^k(\zeta)$ even under the twisted vacuum,
and thus it has an expansion of the form
$W^k(\zeta) \,=\, \sum_{n\in\bZ}\zeta^{-n-k}W^k_n$.
The coefficients form a (linear) basis of the $W_{1+\infty}$ algebra
of central charge $p$.
We will see later in this section,
that $\dphi_0(\zeta)$ is the string field
that creates a loop boundary with $\zeta$.

This $W_{1+\infty}$ currents are simply represented
if we introduce free fermions $c_a(\zeta)$ and $\cb_a(\zeta)$
$(a=0,1,\cdots,p-1)$ through bosonization:
\beqa
  c_a(\zeta)\,=\,K_a\,:e^{-\rphi_a(\zeta)}:\,,~~~~~
  \cb_a(\zeta)\,=\,K_a\,:e^{\rphi_a(\zeta)}:\,.
\eeqa
The cocycle factor $K_a$ is necessary to be introduced
in order to ensure the anticommutation relations with
two different indices ($a\neq b$).
An explicit form may be
$K_a\,=\,(-1)^a\,\prod_{b=0}^{a-1}(-1)^{p_b}$,
where $p_a$ is the momentum of $\dphi_a(\zeta)$.
Thus, $c_a(\zeta)$ and $\cb_a(\zeta)$ $(a=0,1,\cdots,p-1)$
are free fermions satisfying the following OPE and monodromy:
\beqa
  \cb_a(\zeta)\,c_b(\zeta')&\sim&
        \frac{\delta_{a\,b}}{\zeta-\zeta'}\,,~~~~~~~~~~~
  \cb_a(\zeta)\,c_b(\zeta')\,=\,-\,c_b(\zeta')\,\cb_a(\zeta)
        {}~~~(a\neq b) \\
  c_a(\zeta)\,c_b(\zeta')&=&-\,c_b(\zeta')\,c_a(\zeta)\,,~~~~~
  \cb_a(\zeta)\,\cb_b(\zeta')\,=\,-\,\cb_b(\zeta')\,\cb_a(\zeta)\\
  \svac\,c_a(e^{2\pi i}\zeta)&=& \left\{
        \begin{array}{ll}
                -\,\svac\,c_{a+1}(\zeta)~~~~& (0\leq a\leq p-2) \\
                (-1)^{p-1}\svac\,c_0(\zeta)~~~~& (a=p-1)
        \end{array} \right.\\
  \svac\,\cb_a(e^{2\pi i}\zeta)&=& \left\{
        \begin{array}{ll}
                -\,\svac\,\cb_{a+1}(\zeta)~~~~& (0\leq a\leq p-2) \\
                (-1)^{p-1}\svac\,\cb_0(\zeta)~~~~& (a=p-1).
        \end{array} \right.
\eeqa
It is easy to see that the $W_{1+\infty}$ currents can be rewritten
in the following simple form:
\beqa
  W^k(\zeta)&=&\sum_{a=0}^{p-1}:\cb_a(\zeta)
        \del_\zeta^{k-1}c_a(\zeta):\n\\
  {}~&\equiv&\sum_{a=0}^{p-1}\lim_{\zeta'\rightarrow\zeta}
                \del_\zeta^{k-1}\left(\cb_a(\zeta')c_a(\zeta)
                \,-\,\frac{1}{\zeta'-\zeta}\right).
\eeqa

Now that we have necessary machinery,
we can make more explicit statement on the state $\ket{\Phi} \,=\,
\Phi(0)\ket{\rm vac}$
as such that gives the same monodromy with $\ket{\sigma}$
to $\dphih_k(\zeta)$, $\dphi_a(\zeta)$, $c_a(\zeta)$, $\cb(\zeta)$,
and that satisfies the following constraint:\footnote{
We comment that this condition can be restated that for any contour
integral $\oint_C \frac{d\zeta}{2\pi i}\, \zeta^\alpha \, W^k(\zeta)$
($\alpha$: nonnegative integer)
with a closed path $C$ surrounding the origin, where $\Phi(0)$ is
inserted, we can move the path freely in such a way
that it does not surround the origin.
In fact, this observation can be used to calculate various correlation
functions, although we do not describe it in the present paper.}
\beqa
  W^k_n\,\pvac\,=\,0~~(n\geq -k+1).
\eeqa
This constraint is independent of which basis we choose
for the $W_{1+\infty}$ algebra.
In fact, any other basis can be obtained from $W^k(\zeta)$ as
$W^{\prime\,k}(\zeta)\,\equiv\,\sum_n \zeta^{-n-k} \,
W^{\prime\,k}_n \,=\, \sum_{l=0}^{k-1} \beta_l \,
\del_\zeta^l W^{k-l}(\zeta)$ ($\beta_l$: constant),
and thus the same form of constraint also holds
for the $W^{\prime k}_n$'s \cite{fkn2}.

Since $\dphi_a(\zeta) \, = \, :\cb_a(\zeta)c_a(\zeta):$
$(a=1, \cdots, p-1)$ can be obtained from
$\dphi_0(\zeta)$ by letting $\zeta$ go around $\zeta=\infty$
$a$ times, we only have to consider $\dphi_0(\zeta)$.
It is easy to see that the background
$\bra{-B/g} \equiv \svac\exp\left\{ -(1/g)\,
\sum_n B_n \,\alpha_n \right\}$ can be written as
$\svac\exp\{ -(1/g)\,$ $\gint
\frac{d\zeta}{2\pi i}$ $B(\zeta)$ $\dphi_0(\zeta)\}$
with $B(\zeta) \,=\, \sum_{n\geq 1}B_n\, \zeta^{n/p}$.
Note that we here have to make contour integrals $p$ times
around $\zeta=\infty$, since $\dphi_0(\zeta)$ has a nontrivial
monodromy.
We thus define the generating functional for the connected
correlation functions of $\dphi_0(\zeta)$ as
\beqa
  {}F[j(\zeta)]&\equiv&\log\,
        \left\langle -\,\frac{B}{g} \left|\,\,
        :\exp\left\{ \gint \,
        \frac{d\zeta}{2\pi i}\, j(\zeta)\,\dphi_0(\zeta)\,
        \right\}:\,\,
        \right| \Phi\right\rangle \label{2.19}\\
  &=&\log\, \left\langle \sigma \left|\,
        \exp\left\{ -\frac{1}{g}\gint \,
        \frac{d\zeta}{2\pi i}\, B(\zeta)\,\dphi_0(\zeta)\,\right\}
        :\exp\left\{ \gint \,
        \frac{d\zeta}{2\pi i}\, j(\zeta)\,\dphi_0(\zeta)\,
        \right\}:\,
        \right| \Phi\right\rangle. \n
\eeqa

We claim that $F[j(\zeta)]$ is exactly the generating
functional of $w(\zeta_1,\cdots,\zeta_N;\,g)$ in the previous section:
\beqa
  {}F[j(\zeta)]&=&\sum_{N\geq0}\frac{1}{N!}\gint
        \frac{d\zeta_1}{2\pi i}\cdots\frac{d\zeta_N}{2\pi i}\,
        j(\zeta_1)\cdots j(\zeta_N)\,w(\zeta_1,\cdots,\zeta_N;\,g)\n\\
  &=&\sum_{h\geq0}\sum_{N\geq0}\frac{g^{-2+2h+N}}{N!}\,\gint
        \frac{d\zeta_1}{2\pi i}\cdots\frac{d\zeta_N}{2\pi i}\,
        j(\zeta_1)\cdots j(\zeta_N)\,w^{(h)}(\zeta_1,\cdots,\zeta_N).
\eeqa
To prove this, we first note that the symbol $:\,:$ in \eq{2.19}
is the normal ordering that respects the $SL(2,\bC)$ invariant vacuum
$\bra{{\rm vac}}$.
On the other hand, the normal ordering in terms of creation
($\alpha_{-n}$)
and annihilation ($\alpha_{+n}$) operators respects the twisted
vacuum $\svac$.
The difference between these two normal orderings gives rise to
a finite renormalization, and can be calculated explicitly:
\beqa
  :\exp\left\{ \gint \,
        \frac{d\zeta}{2\pi i}\, j(\zeta)\,\dphi_0(\zeta)\,
        \right\}:&=&
  \exp\left\{\,\frac{1}{2}\,\gint \frac{d\zeta_1}{2\pi i}
        \frac{d\zeta_2}{2\pi i}\,N_2(\zeta_1,\,\zeta_2)j(\zeta_1)\,
        j(\zeta_2)\right\}\,\times \\
  &~&\exp\left\{ \gint\,\frac{d\zeta}{2\pi i}
        j(\zeta)\,\dphi_0^{(-)}(\zeta)
        \right\}\,
        \exp\left\{ \gint\,\frac{d\zeta}{2\pi i}
        j(\zeta)\,\dphi_0^{(+)}(\zeta)
        \right\}.\n
\eeqa
Here $\dphi_0^{(\pm)}(\zeta)$ is the part of $\dphi_0(\zeta)$
consisting only of $\alpha_{\pm n}$, respectively,
and $N_2(\zeta_1,\,\zeta_2)$ is given by
\beqa
    N_2(\zeta_1,\,\zeta_2) &=& \frac{1}{p^2\,(\zeta_1\zeta_2)^{1-1/p}\,
        (\zeta_1^{1/p}-\zeta_2^{1/p})^2}
        \,-\,\frac{1}{(\zeta_1-\zeta_2)^2}.
\eeqa
{}Furthermore, picking up the contribution coming from the commutation
between the background term and the source term with negative modes,
we obtain the following expression:
\beqa
  {}F[j(\zeta)]&=&-\frac{1}{g}\,\gint\,\ddz\,N_1(\zeta)\,j(\zeta)\,+\,
        \frac{1}{2}\gint \frac{d\zeta_1}{2\pi i}
        \frac{d\zeta_2}{2\pi i}\,N_2(\zeta_1,\,\zeta_2)j(\zeta_1)\,
        j(\zeta_2) \n\\
  &~&~+\,\log\,
        \left\langle \sigma \left|\,
        \exp\left\{ \gint\,
        \frac{d\zeta}{2\pi i}\,
        \left(j(\zeta)\,-\,\frac{1}{g}B(\zeta)\right)\,
        \dphi_0^{(+)}(\zeta)\,
        \right\}
        \right| \Phi\right\rangle,
\eeqa
where $N_1(\zeta) = \del B(\zeta)$,
and we have used $\svac \dphi_0^{(-)}(\zeta)\,=\, 0$.
Note also that $\svac\dphi_0^{(+)}(\zeta) \,= \, (1/p)\,\sum_{n\geq1}\,
\zeta^{-n/p-1}\, \svac\alpha_n$.
Since the positive-power part $j^{(+)}(\zeta) \equiv \sum_{n\geq1}
j_n\,\zeta^n$ of $j(\zeta)$ only comes up in the bracket,
we finally obtain
\beqa
  \vev{\dphi_0(\zeta_1)\cdots\dphi_0(\zeta_N)}_c&\equiv& \left.
        \frac{\delta}{\delta j(\zeta_1)}\cdots
        \frac{\delta}{\delta j(\zeta_N)}
        \,F[j(\zeta)]\,\right|_{j=0} \n\\
  &=&-\,\frac{1}{g}\,N_1(\zeta_1)\,\delta_{N,\,1}\,
        +\,N_2(\zeta_1,\,\zeta_2)\,\delta_{N,\,2}\\
  &~&~~~+\,\,\frac{1}{p^N}\,\sum_{n_1,\cdots,n_N\geq1}
        \zeta_1^{-n_1/p-1}\cdots\zeta_N^{-n_N/p-1}\,
        \vev{\cO_{n_1}\cdots\cO_{n_N}}_c. \n
\eeqa
The first two terms gives the non-universal terms,
\eq{1.4} and \eq{1.5},
and thus we have established the desired relation:
\beqa
  w(\zeta_1,\cdots,\zeta_N)\,=\,\vev{\dphi_0(\zeta_1)\cdots
        \dphi_0(\zeta_N)}_c.
\eeqa

%%%%%%%%%%%%%%%%%%%%%%%%%%%%%%%%%%%%%%%%%%%%%%%%%%%%%%%%%%%%%%%%%%%
% section 3
%%%%%%%%%%%%%%%%%%%%%%%%%%%%%%%%%%%%%%%%%%%%%%%%%%%%%%%%%%%%%%%%%%%
\section{\protect\large{\protect\bf Soliton Operators and Their Nonperturbative
Effects}}
\cleqn

Since the string fields $\dphi_a(\zeta)$ (with $a=0$ being enough
as explained in the previous section) are composites of
the fermions $c_a(\zeta)$ and $\cb_a(\zeta)$,
we can interpret $c_a(\zeta)$ and $\cb_a(\zeta)$ as
more fundamental dynamical variables in noncritical string field
theories.
If we adopt this point of view, it is quite natural to consider
other bilinears of fermions with vanishing fermion-number:
\beqa
  S_{ab}(\zeta)&\equiv&\cb_a(\zeta)\,c_b(\zeta)~~~(a\neq b)\n\\
  &=&\left\{
        \begin{array}{ll}
          -\,K_a K_b\,e^{\rphi_a(\zeta)-\rphi_b(\zeta)}~~~~&(a<b)\\
          +\,K_a K_b\,e^{\rphi_a(\zeta)-\rphi_b(\zeta)}~~~~&(a>b)\,.
        \end{array}
        \right.
\eeqa
Note that no normal ordering is required since $a\neq b$.
In this section, we show that this is a soliton field
which generates a state with the expectation value of order
$e^{-{\rm const}/g}$ in the weak coupling region ($g\sim0$).

The most important property of this soliton field is
that its commutation with the generators of $W_{1+\infty}$
gives a total derivative:
\beqa
  \left[\, W^k_n ,\, S_{ab}(\zeta)\,\right]\,=\,
        \del_\zeta K^k_{n,\,ab}(\zeta)~~~~(k\geq1,~n\in\bZ),
\eeqa
where $ K^1_{n,\,ab}(\zeta)=0$ and
$K^k_{n,\,ab}(\zeta)= -\,\sum_{l=0}^{k-2}(-1)^l\,
  \del_\zeta^l\left( \zeta^{n+k-1}\,\cb_a(\zeta)\right)\,
  \del_\zeta^{k-l-2} c_b(\zeta)$ $(k\geq 2)$.
Thus, if we introduce the global soliton operator $D_{ab}$
as the contour integral of $S_{ab}(\zeta)$
\beqa
  D_{ab} \,=\,\gint \frac{d\zeta}{2\pi i}S_{ab}(\zeta),
\eeqa
we see that $D_{ab}$ commutes with the generators $W^k_n$:
\beqa
  \left[\, W^k_n,\, D_{ab}\,\right]&=&0~~~~
        (k\geq1,~n\in\bZ). \label{3.6}
\eeqa
Note that the contour must surround the origin $p$ times,
since the $S_{ab}(\zeta)$ has a monodromy around that point:
$S_{ab}(e^{2\pi i}\zeta)\ket{\Phi} \,=
\,S_{[a+1]\,[b+1]}(\zeta) \ket{\Phi}$.
Eq.\ \eq{3.6} implies that
if a state $\ket{\Phi}$ satisfies the Schwinger-Dyson equations
$W^k_n\,\ket{\Phi}\,=\,0$ $(k\geq 1,~ n\geq -k+1)$,
then $\ket{D_{ab}}\, \equiv \,D_{ab}\,\ket{\Phi}$ also satisfies
the same equations
$W^k_n\,\ket{D_{ab}}\,= \,0$.
In this sense, we have found that the Schwinger-Dyson equations
do not determine the vacuum uniquely.

Interpreting this state $\ket{D_{ab}}$ as 1-soliton
background,\footnote{
Multi-instanton background will be obtained by
repeatedly inserting the operators $D_{ab}$.}
we can calculate its nonperturbative effect explicitly
for the unitary case $(p,q)=(p,p+1)$.
This will be defined by the ratio
\beqa
  A_{ab}\,\equiv\,\frac{
        \left\langle\left.\,-\,\frac{B}{g}\,\right|\,D_{ab}\,\right\rangle
        }{
        \left\langle\left.\,-\,\frac{B}{g}\,\right|\,\Phi\,\right\rangle
        }
  \,\sim\,\vev{\gint \frac{d\zeta}{2\pi i}\,
        e^{\rphi_a(\zeta)-\rphi_b(\zeta)}}.
\eeqa
Here we have neglected the irrelevant contribution from the cocycles.
In the weak coupling region, the leading contribution comes from
the disk amplitude:
\beqa
  A_{ab}\,=\,\gint\ddz e^{(1/g)\Gamma_{ab}(\zeta)+O(g^0)},
\eeqa
where the ``effective action'' $\Gamma_{ab}(\zeta)$:
\beqa
  \Gamma_{ab}(\zeta)\,=\,\vev{\rphi_a(\zeta)}^{(0)} \,-\,
        \vev{\rphi_b(\zeta)}^{(0)}
\eeqa
can be read off from the disk amplitude
$w^{(0)}(\zeta)=\vev{\dphi_0(\zeta)}^{(0)}$.
In fact, $w^{(0)}(\zeta)$ was calculated in ref.\ \cite{mss},
and has the following form:\footnote{
The present normalization corresponds to the background
$B_1=t\, (>0)$, $B_{2p+1}=-4p/(p+1)(2p+1)$
and $B_n=0~(n\neq 1,\,2p+1)$.
$B_{2p+1}$ must be negative when $t>0$ in order for
$w^{(0)}(\zeta)$ to have a cut only on the negative real axis.}
\beqa
  w^{(0)}(\zeta)&=&\vev{\dphi_0(\zeta)}^{(0)}\n\\
  {}~&=&\beta_p\,\left[\, \left(\zeta+\sqrt{\zeta^2-t}\right)^r \,+\,
         \left(\zeta-\sqrt{\zeta^2-t}\right)^r \,\right],
\eeqa
where $r=q/p=(p+1)/p$ and $\beta_p=2^{(p-1)/p}/(p+1)$.
Since $\vev{\dphi_0(\zeta)}^{(0)}=
\sum_{k=0}^{p-1}\vev{\dphih_k(\zeta)}^{(0)}$,
and $\vev{\dphih_k(\zeta)}^{(0)}$ behaves as
$(1/p)\sum_{n\equiv k\, ({\rm mod}\,p)} \zeta^{-n/p-1}\vev{\cO_n}^{(0)}$
around $\zeta=\infty$ up to non-universal part,
we find that
\beqa
  \vev{\dphih_1(\zeta)}^{(0)}&=&
        \beta_p\,\left(\zeta-\sqrt{\zeta^2-t}\right)^r \n\\
  \vev{\dphih_{p-1}(\zeta)}^{(0)}&=&
        \beta_p\,\left(\zeta+\sqrt{\zeta^2+t}\right)^r \n\\
  \vev{\dphih_k(\zeta)}^{(0)}&=&0~~~(k\neq0,~p-1).
\eeqa
We can in turn calculate $\vev{\dphi_a(\zeta)}^{(0)}$ as
\beqa
  \vev{\dphi_a(\zeta)}^{(0)}&=&\sum_{k=0}^{p-1}\omega^{-ka}
        \vev{\dphih_k(\zeta)}^{(0)}\n\\
  &=&\beta_p\,\left[\,
        \omega^{-a}\,\left(\zeta-\sqrt{\zeta^2-t}\right)^r\,+\,
        \omega^a\,\left(\zeta+\sqrt{\zeta^2-t}\right)^r\,\right],
\eeqa
and $\Gamma_{ab}$ is obtained by integrating
$\vev{\dphi_a(\zeta)}^{(0)}-\vev{\dphi_b(\zeta)}^{(0)}$.

We here notice that for $\dphi_a(\zeta)$ with $a\neq 0$,
there is an extra cut of second order
along $-\sqrt{t}\leq \zeta\leq\sqrt{t}$,
in addition to the ``physical'' cut of order $p$
on the negative real $\zeta$ axis
which $\dphi_0(\zeta)$ has originally.
It might seem strange that
there appears a function which gives a cut
with a leg on the positive real axis.
However, in the situation we consider,
such function is always integrated on some contour,
and thus this kind of cut will never be observed
in final results.
To cope with this ``unphysical'' cut,
we introduce a new coordinate $s$ which is defined by
\beqa
  s(\zeta)\,=\,\frac{1}{\sqrt{t}}\left(\zeta+\sqrt{\zeta^2-t}\right).
\eeqa
Here $s(\zeta)$ maps two sheets of the $\zeta$-plane with cut,
onto the outside and the inside of the unit circle in the $s$-plane,
respectively.
Note that the original physical cut is now mapped to
$-\infty<s<0$, and also that
two points on different sheets with the same $\zeta$ coordinate,
are related by the map $s\mapsto 1/s$ in the $s$-plane.

Making use of this $s$-coordinate, the first derivative of
the effective action $\Gamma_{ab}$ is found to be
\beqa
  \del_s\Gamma_{ab}(s)&\equiv&\frac{d\zeta}{d s}\,
        \left(\,\vev{\dphi_a(\zeta)}^{(0)} \,-\,
        \vev{\dphi_b(\zeta)}^{(0)}\,\right)\n\\
  &=&\frac{1-s^{-2}}{2}\,\beta_p\,t^{(r+1)/2}\,
        \left[\,(\omega^a-\omega^b)\,s^r\,+\,
        (\omega^{-a}-\omega^{-b})\,s^{-r}\,\right].\label{3.14}
\eeqa
By integrating this with respect to $s$,
$\Gamma_{ab}(s)$ is easily obtained to be
\beqa
  \Gamma_{ab}(s)&=&\gamma_p\,\left[\, \frac{1}{r+1}\left\{
        (\omega^a-\omega^b)s^{r+1}+
        (\omega^{-a}-\omega^{-b})s^{-(r+1)}\right\}
        \right.\n\\
  &~&~~~~~-\,\frac{1}{r-1}\left.\left\{
        (\omega^a-\omega^b)s^{r-1}+
        (\omega^{-a}-\omega^{-b})s^{-(r-1)}\right\}
        \,\right],
\eeqa
where $\gamma_p=2^{-1/p}\,t^{(2p+1)/2p} /(p+1)$.

Saddle points $s_0$ of $\Gamma_{ab}$ for $g \rightarrow +\,0$
are obtained from the equation $\del_s\Gamma_{ab}(s) = 0$.
Among them, the points $s_0=\pm1$, coming from
$d\zeta/d s = \sqrt{t}\,(1-s^{-2})/2 = 0$ in \eq{3.14}, do not
contribute to the contour integral due to the Jacobian factor:
$\gint\ddz=\gint\dds\,\frac{d\zeta}{d s}$.
Thus, it is enough to consider the saddle points
coming only from the latter factor in \eq{3.14}, and we find
$s_0^{2r} = -(\omega^{-a}-\omega^{-b}) / (\omega^a-\omega^b) =
\omega^{-a-b} = \omega^{np-a-b}$, namely,
\beqa
  s_0\,=\,e^{(np-a-b)\pi i/(p+1)}.
\eeqa
Here the integer $n$ was introduced to take into account
on which (physical) sheet $s_0$ lives.
Then $\Gamma_{ab}(s_0)$ and $\del_s^2\Gamma_{ab}(s_0)$ are
easily evaluated to be
\beqa
  \Gamma_{ab}(s_0)&=&\frac{8p}{2^{1/p}\cdot(2p+1)}\,t^{\frac{2p+1}{2p}}
        \sin\left(\frac{a+b+n}{p+1}\pi\right)\,
        \sin\left(\frac{a-b}{p}\pi\right)\n\\
  \frac{d^2}{ds^2}\Gamma_{ab}(s_0)&=&
        \frac{2p+1}{p^2}\,s_0^{-2}\,\Gamma_{ab}(s_0). \label{3.17}
\eeqa

In order to investigate the nonperturbative effects of stable solitons
which give small and finite contributions in the weak coupling limit,
$g\rightarrow +\,0$, we choose a contour along which
${\rm Re}\left(\Gamma_{ab}(s)\right)$ only takes negative values.
The maximum contribution to $A_{ab}$ can then be picked up
from saddle points if the contour passes these points
in the direction of steepest descent, along which
$\del_s^2\Gamma_{ab}(s_0)(s-s_0)^2$ is negative.
According to \eq{3.17}, such direction is given by
${\arg}((s-s_0)^2/{s_0}^2)=0$ when $\Gamma_{ab}(s_0)<0$,
namely, the path goes through the unit circle perpendicularly
at $s=s_0$.

As an example, we first consider $A_{01}$
in the $p=2$ (pure gravity) case.
Then, the saddle points $s_0$ satisfying $\Gamma_{01}(s_0)<0$
are only on the first physical sheet $(-\pi\leq {\rm arg}(s)<\pi)$:
$s_0=e^{\pm\pi i/3}$.
As a path of integration passing these saddle points on the first
physical sheet, we find the one depicted as $C_1$ in Fig.\ 1.
On the other hand, since a path on the second physical sheet
$(\pi\leq {\rm arg}(s)<3\pi)$
must be also chosen so that the $\Gamma_{01}(s)$ only takes negative
values on it,
the path turns out to go around
the point of infinity in the region $9\pi/5< \arg(s) < 11\pi/5$.
Thus, the contribution to $A_{01}$ comes only from the integration
around the saddle points on the first physical sheet,
and we get
$A_{01} \sim g^{1/2}\, t^{-1/8}\,
e^{-4\sqrt{6}t^{5/4}/5g}$ from \eq{3.17}.
The exponent in the expression coincides with the value
\cite{bk-ds-gm,s,d2}
obtained in the nonperturbative analysis
of the string equation (Painlev\'{e} equation):
$4u^2 + (2g^2/3)\del_t^2u = t$, where $u = \vev{\cO_1\cO_1}_c$.
A similar consideration can be applied to $A_{10}$,
giving the same value with $A_{01}$,
but now the contribution comes from the second physical sheet
with $s_0 = e^{5\pi i/3},~e^{7\pi i/3}$.

\begin{figure}
\begin{center}
\leavevmode
\epsfbox{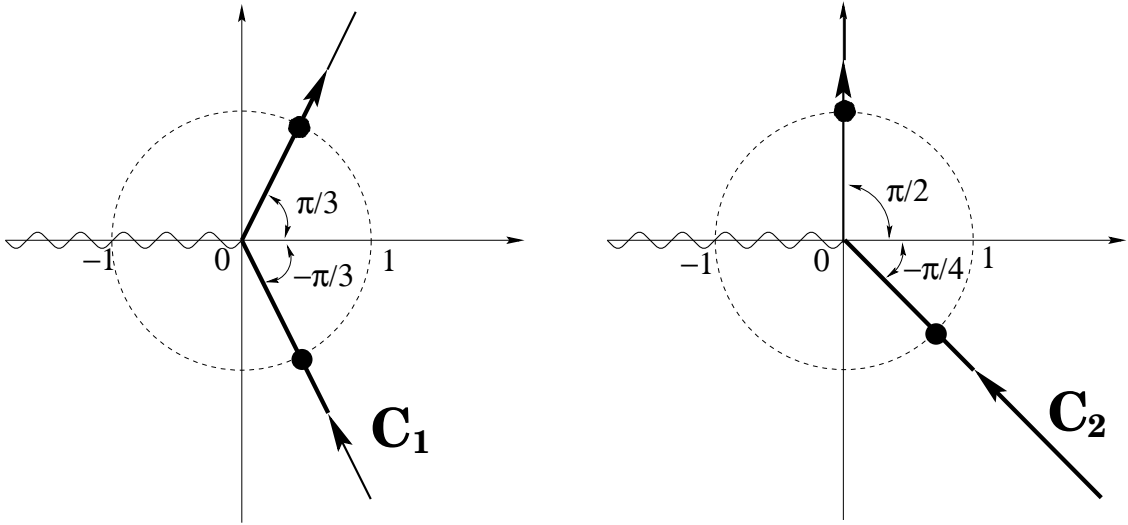}
\end{center}
\caption{The heavy solid line in the left (right) figure denotes
a path of integration $C_1$ ($C_2$) on the first physical sheet
for $p=2$ ($p=3$). The blobs describe the saddle points,
and the physical cuts are on the negative real axis.}
\end{figure}

We finally consider the $p=3$ (Ising) case.
Here we have three physical sheets.
{}For $A_{01}$,
there are two saddle points satisfying $\Gamma_{01}(s_0)<0$ on the
first physical sheet $(-\pi\leq {\rm arg}(s)<\pi)$:
$s_0=e^{-\pi i/4}$ and $s_0=e^{\pi i/2}$.
The path of integration passing these saddle points can be
taken as $C_2$ depicted in Fig.\ 1.
The saddle point at  $s_0=e^{-\pi i/4}$ gives the leading
nonperturbative effect
$A_{01} \sim g^{1/2}\, t^{-1/12}\,
e^{-6\sqrt{6}t^{7/6}/2^{1/3}\cdot7g}$.
A smaller contribution to $A_{01}$ also appears
from the saddle point $s_0=e^{\pi i/2}$
as $\sim g^{1/2}\, t^{-1/12}\,
e^{-12\sqrt{3}t^{7/6}/2^{1/3}\cdot7g}$.
Paths on the second physical sheet
$(\pi\leq {\rm arg}(s)<3\pi)$ and the third one
$(3\pi\leq {\rm arg}(s)<5\pi)$
can be taken as going around the point of infinity in the
region where ${\rm Re}(\Gamma_{01}(s)) < 0$,
giving no contribution to $A_{01}$.
The values of the exponent in $A_{01}$ coming from the saddle points
%$s_0=e^{-\pi i/4}$ and $s_0=e^{\pi i/2}$
agree with those
appearing as the nonperturbative effects in the string equation
$4 u^3 + (3g^2/2)(\del_tu)^2
+ 3g^2 u \del_t^2u + (g^4/6) \del_t^4u = - t$
where $u$ is again $\vev{\cO_1\cO_1}_c$.
A similar consideration can be applied to other $A_{ab}$, giving
the same values with $A_{01}$, but, for example, the contributions to
$A_{12}$ and $A_{20}$ come from the saddle points on the third
and the second physical sheet, respectively.

%%%%%%%%%%%%%%%%%%%%%%%%%%%%%%%%%%%%%%%%%%%%%%%%%%%%%%%%%%%%%%%
% section 4
%%%%%%%%%%%%%%%%%%%%%%%%%%%%%%%%%%%%%%%%%%%%%%%%%%%%%%%%%%%%%%%
\section{\protect\large{\protect\bf Conclusions}}
\cleqn

In this paper we first rewrite the Schwinger-Dyson equations
in terms of free fermions, $c_a(\zeta),~\cb_a(\zeta)$,
with twist operator $\sigma(\zeta)$.
We then find that the Schwinger-Dyson equations allow solutions
with soliton backgrounds
constructed from the soliton fields
$S_{ab}$, which are composites of the fermions:
$S_{ab}(\zeta)=\cb_a(\zeta)c_b(\zeta)$ $(a\neq b)$.
Since elementary strings (macroscopic loops) are also given
as their composites, $\dphi_a(\zeta) = {:\cb_a(\zeta)c_a(\zeta):}$,
we suggest that these fermions should be regarded
as the fundamental dynamical variables in noncritical string theory.
We further evaluate the contributions from the stable solitons,
and show that they are consistent with the nonperturbative analysis
of string equations.

We, however, have the impression that the formalism presented here
is still not satisfactory, because we cannot determine
the weights in the summation over multi-instantons,
while they seem to be fixed according to the analysis of string equations.
We expect that these weights will be determined automatically
once the actions for these fermions are constructed.
In fact, actions would be constructed for such fundamental
dynamical variables,
while it seems almost impossible to construct
simple string-field actions only with elementary string fields
(macroscopic loops).
The investigation in this direction is now in progress,
and will be reported elsewhere.

%%%%%%%%%%%%%%%%%%%%%%%%%%%%%%%%%%%%%%%%%%%%%%%%%%%%%%%%%%%%%%%%%%%%%
% Acknowledgment
%%%%%%%%%%%%%%%%%%%%%%%%%%%%%%%%%%%%%%%%%%%%%%%%%%%%%%%%%%%%%%%%%%%%%
\section*{\protect\large{\protect\bf Acknowledgment}}
We would like to thank N.\ Ishibashi, A.\ Ishikawa, K.\ Itoh,
H.\ Kanno, H.\ Kawai, T.\ Kawano, T.\ Kugo and M.\ Ninomiya
for useful discussions.
We are also grateful to the organizers of Kashikojima workshop
where an essential part of the present work was developed.
This work is supported in part by the Grant-in-Aid for Scientific
Research from the Ministry of Education, Science and Culture.

%%%%%%%%%%%%%%%%%%%%%%%%%%%%%%%%%%%%%%%%%%%%%%%%%%%%%%%%%%%%%%%%%%%%%
% References
%%%%%%%%%%%%%%%%%%%%%%%%%%%%%%%%%%%%%%%%%%%%%%%%%%%%%%%%%%%%%%%%%%%%%


\begin{thebibliography}{99}

\bibitem{m-soliton}
E.\ Witten, {\em Nucl. Phys.} {\bf B443} (1995) 85;\\
M.\ Duff, R.\ Khuri and J.\ Lu, {\em Phys. Rept.} {\bf 259} (1995) 213;\\
A.\ Sen, {\em Nucl. Phys.} {\bf B450} (1995) 103;\\
J.\ Harvey and A.\ Strominger, {\em Nucl. Phys.} {\bf B449} (1995) 535

\bibitem{s}
S.\ Shenker, in {\em Cargese 1990, Proceedings: Random Surfaces and
Quantum Gravity} (1990) 191

\bibitem{p1}
J.\ Polchinski, Phys. Rev. Lett. {\bf 75} (1995) 4724;\\
J.\ Dai, R.\ Leigh and J.\ Polchinski, {\em Mod. Phys. Lett.}
{\bf A4} (1989) 2073

\bibitem{p2-g}
J.\ Polchinski, {\em Phys. Rev.} {\bf D50} (1994) 6041;\\
M.\ Green, {\em Phys. Lett.} {\bf B354} (1995) 271

\bibitem{bk-ds-gm}
E.\ Br\'{e}zin, and V.\ Kazakov, {\em Phys. Lett.}
{\bf B236} (1990) 144;\\
M.\ Douglas and S.\ Shenker, {\em Nucl. Phys.}
{\bf B335} (1990) 635;\\
D.\ Gross and A.\ Migdal, {\em Phys. Rev. Lett.} {\bf 64} (1990) 127;
Nucl. Phys. {\bf B340} (1990) 333

\bibitem{d0}
{}F.\ David, {\em Mod. Phys. Lett.} {\bf A5} (1990) 1019

\bibitem{fkn1}
M.\ Fukuma, H.\ Kawai and R.\ Nakayama,
{\em Int. J. Mod. Phys.} {\bf A6} (1991) 1385

\bibitem{dvv}
R.\ Dijkgraaf, E.\ Verlinde and H.\ Verlinde,
{\em Nucl. Phys.} {\bf B348} (1991) 435

\bibitem{bdss}
T.\ Banks, M.\ Douglas, N.\ Seiberg and S.\ Shenker,
{\em Phys. Lett.} {\bf B238} (1990) 279

\bibitem{pk-kpz-ddk}
A. Polyakov, {\em Mod. Phys. Lett.} {\bf A2} (1987) 899;\\
V. Knizhnik, A. Polyakov and A. Zamolodchikov,
{\em Mod. Phys. Lett.} {\bf A3} (1988) 819;\\
{}F. David, {\em Mod. Phys. Lett.} {\bf A3} (1988) 1651;\\
J. Distler and H. Kawai, {\em Nucl. Phys.} {\bf B321} (1989) 509

\bibitem{fkn2}
M.\ Fukuma, H.\ Kawai and R.\ Nakayama,
{\em Commun. Math. Phys.} {\bf 143} (1992) 371

\bibitem{md}
M.\ Douglas, {\em Phys. Lett.} {\bf B238} (1990) 176

\bibitem{fkn3}
M.\ Fukuma, H.\ Kawai and R.\ Nakayama,
{\em Commun. Math. Phys.} {\bf 148} (1992) 101

\bibitem{winf}
C.\ Pope, L.\ Romans and X.\ Shen, {\em Phys. Lett.} {\bf B242} (1990) 401;\\
V.\ Kac and A.\ Radul, {\em Commun. Math. Phys.} {\bf 157} (1993) 429;\\
H.\ Awata, M.\ Fukuma, Y.\ Matsuo and S.\ Odake,
{\em Prog. Theor. Phys. Suppl.} {\bf 118} (1995) 343 
and references therein

\bibitem{gn}
E.\ Gava and K.\ Narain, {\em Phys. Lett.} {\bf B263} (1991) 213

\bibitem{mss}
G.\ Moore, N.\ Seiberg and M.\ Staudacher,
{\em Nucl. Phys.} {\bf B362} (1991) 665

\bibitem{iikmns}
M.\ Ikehara, N.\ Ishibashi, H.\ Kawai, T.\ Mogami, R.\ Nakayama
and N.\ Sasakura,
{\em Phys. Rev.} {\bf D50} (1994) 7467

\bibitem{dfms-br}
L.\ Dixon, D.\ Friedan, E.\ Martinec and S.\ Shenker,
{\em Nucl. Phys.} {\bf B282} (1987) 13;\\
M.\ Bershadsky and A.\ Radul,
{\em Int. J. Mod. Phys.} {\bf A2} (1987) 165

\bibitem{d2}
{}F.\ David,
in {\em Two Dimensional Quantum Gravity and Random Surfaces},
World Scientific (1992) 125

\end{thebibliography}
\end{document}